\documentclass{emulateapj}

\usepackage{epsfig}

\def\deg{^{\circ}}
\def\P3hat{{\mathaccent 94 P}_3}

\def\A{{\bf A}~}
\def\B{{\bf B}~}
\def\An{{\bf A}}
\def\Bn{{\bf B}}
\def\pA{pulsar {\bf A}~}
\def\pB{pulsar {\bf B}~}
\def\pAn{pulsar {\bf A}}
\def\pBn{pulsar {\bf B}}
\def\snr{{\tt S/N~}}
\def\phorb{{\phi_{\rm orb}}}

\shorttitle{Morphological properties of J0737--3039 pulsars}

\shortauthors{R. Ramachandran et al.}

\begin{document}

\title{Fluctuation and morphological properties of the pulsars in J0737--3039 system}

\author{R. Ramachandran, D. C. Backer, P. Demorest} \affil{Department
of Astronomy, University of California, Berkeley, CA 94720-3411, USA}

\author{S. M. Ransom$^{1}$, V. M. Kaspi$^{1,2}$}

\affil{Department of Physics, McGill
University, Montreal, QC H3A 2T8, Canada} \altaffiltext{1}{Center for
Space Research, Massachusetts Institute of Technology, Cambridge, MA
02139} \altaffiltext{2}{Canada Research Chair, Steacie Fellow, CIAR
Fellow}

\begin{abstract}
We describe the morphological and fluctuation properties of the
pulsars in the double neutron star system, PSR J0737--3039. Pulsar \B
is seen in almost all orbital phases, except in the range of $\sim
6\deg$ to $65\deg$. This may be interpreted as an {\it eclipse} of
\pBn's signal by its own magnetopause region produced by interaction
with \pAn's relativistic wind. No modulation of the emission of \pB is
found at the period of \pAn. This places a constraint on the models
that propose that \pAn's beamed radiation is directly responsible for
\pBn's emission. Modulation index values indicate that the pulse to
pulse variations in the two objects are mostly intrinsic. Pulsar A
shows significant differential modulation index within its pulse
profile.
\end{abstract}


\section{Introduction}
\label{sec-intro}
Recent discovery of the double pulsar system, reported by Burgay et
al. (2003) and Lyne et al. (2004) is one of the most important
discoveries in the history of pulsar astronomy. This high inclination system,
which consists of a short period recycled pulsar and an ordinary slow
pulsar revolving with an orbital period 2.45 hrs,
gives us almost an ideal situation to probe many fundamental properties
of neutron star astrophysics.

The distance between the two pulsars is so small (2.8 light sec), that the
light cylinder volume (magnetosphere of a solitary pulsar) of PSR
J0737--3039B (hereafter \Bn) subtends about 20 degrees from PSR
J0737--3039A (hereafter \An).

It is unclear whether the measured value of $\dot{P}$ of \pB is
intrinsic to the pulsar, given the mechanical luminosity ($\dot{E}$)
of \pAn. It is conceivable that the relativistic wind from \pA
influences the characteristics of \pBn.  In spite of this, \pB
continues to emit radio radiation, clearly demonstrating the
robustness of the fundamental radio emission mechanism.  According to
Lyne et al. (2004), this system may well have indicated that the seed
acceleration responsible for this emission comes from deep inside the
magnetosphere, clearly ruling out outer magnetospheric emission
models.

The measured flux of \pB significantly changes as a
function of orbital phase\footnote{Orbital phase ($\phorb$) is given
as $(\omega+\eta)$, where $\omega$ and $\eta$ are the position angle
of periastron and the ``true'' anomaly.} (Lyne et al. 2004). 

Models have been proposed to suggest that the wind from \pA creates a
magnetopause region around \pB due to the interaction with \pBn's
magnetosphere (Arons et al. 2004). In this model, the synchrotron
absorption optical
depth of this
magnetopause region is the primary reason for the observed eclipse of
\pAn, and possibly for the light curve of \pBn.

In this work, we present some of the basic morphological and
fluctuation properties of these two pulsars. Section \ref{sec-flux}
describes the properties of average pulses of \pA and \pBn. The
fluctuation properties of the two pulsars are described in Section
\ref{sec-fluct}. Implications of our results are discussed in Section
\ref{sec-discuss}.

\section{Observations}
\label{sec-obsvn}
Observations were performed at the Green Bank Telescope (GBT) on 2003
December 11, 19, 23, \& 24, and 2004 January 1. With the BCPM
(Berkeley-Caltech Pulsar Machine; see e.g., Camilo et al. 2002), we
observed at 430, 800, 1400 \& 2200 MHz frequencies to record total
power (Stokes-I parameter) at a time resolution of 72$\mu$sec. The
bandwidths for the first two bands were 48 MHz (96 channels of 0.5 MHz
each), and the last two were 96 MHz (96 channels of 1 MHz each).
At 430 and 800 MHz, we also used the Spigot card with a
bandwidth of 50 MHz.

The data was corrected for interstellar dispersion as part of our
offline processing. We then added all channels together to produce a
single time series. We used the ephemeris given by Burgay et
al. (2003) for analysing the data from December 11th, 2003, and
produced an improved ephemeris for \pAn. This was then used for
generating an ephemeris for \pB for all our subsequent observations.

\section{Profile characteristics and light curves}
\label{sec-flux}

The two pulsars in this binary system show entirely different
characteristics in their average properties. Apart from an eclipse of
$\sim$30 sec duration, \pA shows no significant variation in its
light curve\footnote{Observed flux as a function of time}, whereas \pB
exhibits a spectacular orbital phase dependent variation in its light
curve. Pulsar \A shows significant average profile variation as a
function of radio frequency which is seen in many other MSPs. Pulsar
\B shows significant profile variation not only as a function of
frequency, but also as a function of orbital phase. We summarise all
these properties in this section.

\subsection{Average profiles of \pAn}
\label{sec-profa}
\begin{figure}
\begin{center}
\epsfig{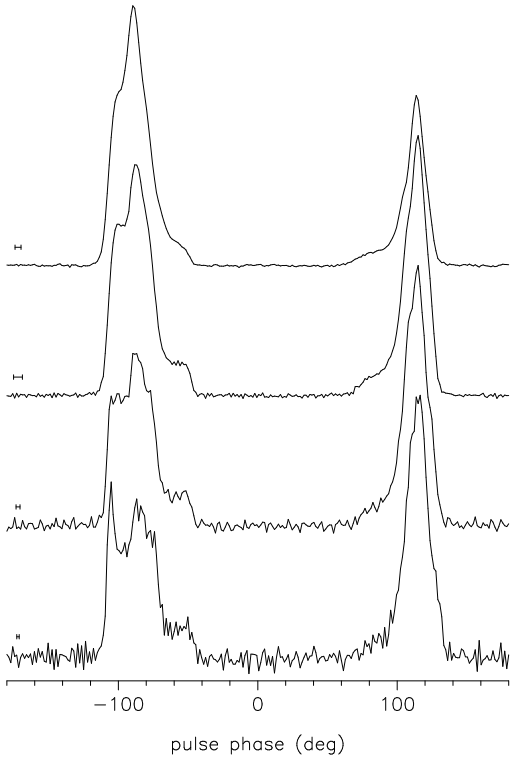}
\caption{Average profiles of \pA at (top to bottom) 427, 820, 1400 \&
2200 MHz. The profile at 427 MHz is from the SPIGOT card, and the rest
are from BCPM. The effective time resolution for each of the profiles
is indicated as a horizontal error bar. See text for details.}
\label{fig:profA}
\end{center}
\end{figure}

Pulsar \A does not show any significant flux variation as a function
of orbital phase outside of the eclipse region at its superior
conjunction, whose characteristics are described by Lyne et al. (2004)
and Kaspi et al. (2004). However, \pA shows considerable change in its
average profiles as a function of radio frequency. We observed this
pulsar at four different frequency bands, namely 427, 820, 1400 \&
2200 MHz. Average profiles constructed from these data sets are given
in Figure \ref{fig:profA}.

As Burgay et al. (2003) \& Lyne et al. (2004) reported, we have a
two-featured profile. Although these features appear like the classic
main pulse and interpulse of an orthogonal rotator, the most favored
configuration is likely an almost aligned rotation and magnetic axes,
where our sight line cuts the emission cone in two different sides of
the cone (Demorest et al. 2004). At the lowest frequency, the first
feature is dominant, with less pronounced structures. However, at all
the other frequencies, the second feature is dominant. Also, the first
feature develops complex component structures. Although this is quite
strikingly different from the behavior of ordinary long period
pulsars, it is by no means uncommon among the shorter period pulsar
population.

The amplitude reversal of components as a function of frequency has
been noted among several short period pulsars (millisecond pulsars).
For instance, PSRs J0751+1807 \& J2145--0750 are known to exhibit such
reversal (Kramer et al. 1999). PSR J1022--1001 exhibits even a more
complex evolution, where the component strengths reverse twice, while
going from 400 MHz to a few GHz (Ramachandran \& Kramer 2003). On the
other hand, there are also counter examples like J0621+1002 \&
J1744--1134, where the pulsar exhibits little change in the profile
structure.

We estimated the width and separation of the two features at all the
frequency bands, and they do not seem to exhibit significant change.
As it turns out, average profiles in all four bands could be fitted
with seven gaussian components (4 for first feature, and 3 for the
second). In order to quantify the profile span in longitude
completely, we measured the width between the ``base'' (corresponds to
approximately 1/50th of the peak value) of the raising and the
trailing edge. This turned out to be 66$\deg$ and 60$\deg$ for the two
features in all frequencies, and these values coincided within
2-3$\deg$. Similarly, the separation of the outer edges of the two
features was $\sim$240$\deg$, and the separation of inner edges was
$\sim$114$\deg$.

Several investigators have studied the variation of pulse width and
component separation as a function of radio frequency. The classical
`conal-single' pulsars are known to exhibit a power law behavior,
where the width of the pulse depends on radio frequency as $\nu^{-b}$,
where $b$ is in the range of $\sim 0.1-0.3$ (Rankin 1983). Hankins \&
Fowler (1986) studied the frequency dependence of separation between
the main and the interpulse of eight pulsars with interpulses. They
find that although the main and the interpulse show considerable
frequency dependence in their shape, the separation between them is
frequency independent. Although this lack of frequency dependence
tends to favor magnetic and rotation axes being almost orthogonal to
each other (observed emission coming from both poles, with main pulse
from one pole and the interpulse from the other pole: {\it two-pole
model}), with their polarization observations, they point out that
they favor the geometry where the two axes are almost aligned ({\it
single-pole model}).

However, for recycled pulsars, the observed frequency dependence is
more complex. The separation between the two features of PSR B1259--63
changes from $\sim$160$\deg$ to 139$\deg$ while going from 660 MHz to
1.5 GHz. Then, at 4.7 GHz, the separation is 107$\deg$, and it stays
constant above this frequency (Manchester \& Johnston 1995). As
Manchester \& Johnston state, the most favored model for this pulsar
is the single-pole model.

In the case of PSR J1022+1001, although the strength of the components
evolve significantly as a function of radio frequency, and the profile
itself shows significant instability over time scales of several
minutes, the location of components and the width of the profile
remains almost constant over a wide range of 328 MHz to about 4.8 GHz
(Ramachandran \& Kramer 2003). Kramer et al. (1999) present a very
good summary of recycled pulsar profiles and their frequency
evolution. In fact, their fig. 16 demonstrates clearly that the
observed $b$ values for recycled pulsars are very close to zero, with
a few exceptions.

Therefore, one can conclude from the behavior of average pulses of \pA
as a function of radio frequency that it is no different from the
general behavior of short period pulsar population. The lack of
frequency dependence of either the pulse width or the main and the
interpulse separation cannot be considered as a strong evidence for
the two-pole model.

\subsection{Light curves of \pBn}
\label{sec-lightcurve}
\begin{figure*}[t]
\begin{center}
\epsfig{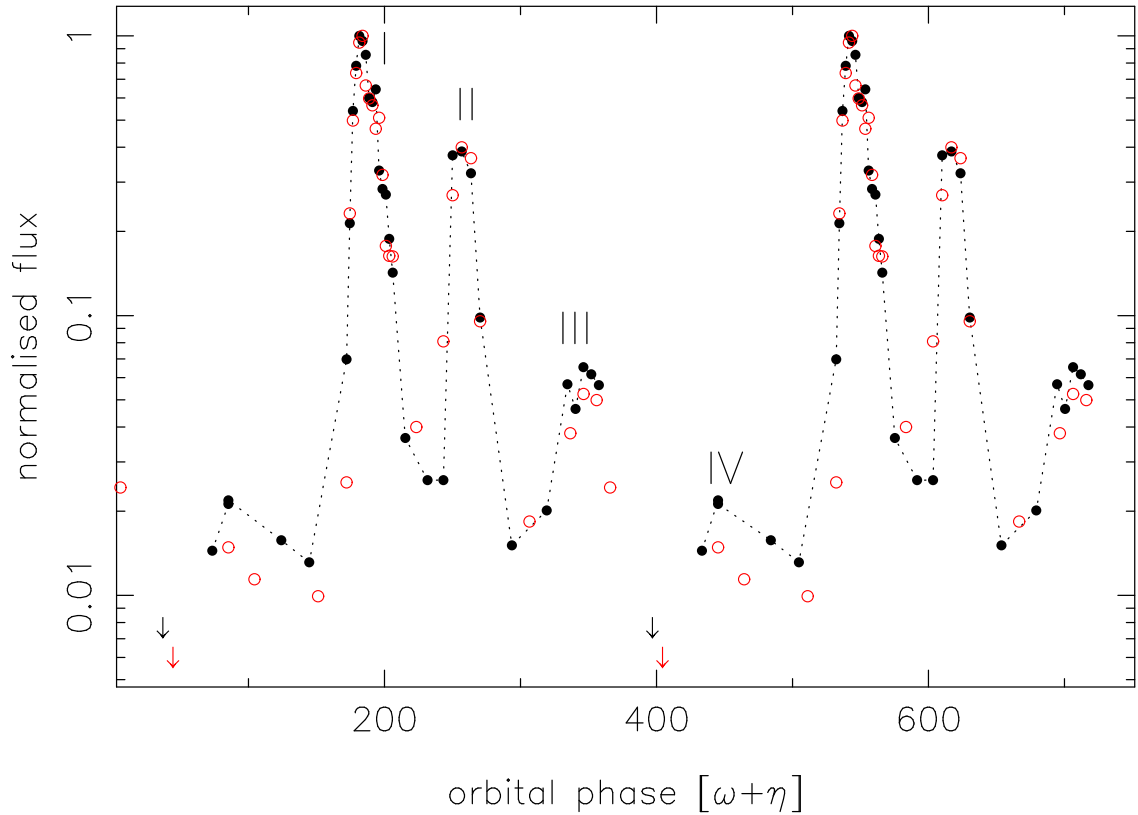}
\caption[]{Light curve of \pB at 820 MHz (filled circles) and 1400 MHz
(open circles). The four bright windows are indicated as I, II, III \&
IV. The dotted line has been drawn just for clarity. See text for
details.}
\label{fig:lc}
\end{center}
\end{figure*}

As Lyne et al. (2004) indicate, \pBn's flux varies systematically as a
function of orbital phase. In some phases, the pulsar is so bright
that single pulses are detectable. However, in most of the orbital
phase range, the emission is faint. Figure \ref{fig:lc} shows that the
pulsar is prominent in four orbital phase ranges (windows I to
IV). For clarity, we have given the relative flux in logarithmic units
and have repeated the light curve for two orbital periods. The
inferior conjunction of \pB (when \B is in front of \An) occurs when
$\phorb = 270\deg$. 

There are several features here that are noteworthy. Each sample in
Figure \ref{fig:lc} is produced by finding the area of the pulse after
averaging some number of pulses. Owing to \snr considerations, we have
taken unequal phase intervals for defining each sample in the
plot. The shortest averaging length that we have taken is when the
pulsar is very bright ($\phorb\sim180\deg$) where we have taken an
average of 20 pulses. The longest length is when the pulsar flux was
the weakest ($\phorb\sim 400\deg$), where we have considered 700
pulses to find an upper limit to the average pulse flux.

It is worth emphasizing one important aspect of this light curve. In
the orbital phase range of $\sim6\deg$ to 65$\deg$ (modulo 360$\deg$
in Figure \ref{fig:lc}), it appears that the pulsar signal exhibits
what may be considered as an ``eclipse''. We did not detect
significant flux in this range at all. Although we cannot define the
boundary of this ``eclipse'' region clearly due to signal to noise
ratio considerations, the boundary appears to be very sharp. The
physical mechanisms responsible for this behavior will be discussed in
detail elsewhere (Arons et al. 2004).

Pulsar profiles reach a stable state only after averaging some
thousand pulses (Helfand et al. 1975). Therefore, it is likely that
our flux estimation in the phase ranges where we have taken very short
averaging are somewhat in error. However, our emphasis here is on the
overall systematic relative variations, and not on the absolute flux
itself. As we will see in the next section, this behavior is not
restricted to only the light curve, but also to the detailed profile
structure.

There is some hint of chromaticity in the light curve during
windows III and IV, where we observed lower flux at 1400 MHz than at
800 MHz. However, when the pulsar is bright, the light curve seems to
be achromatic. A possible reason for the observed chromaticity in
windows III and IV may be due to the line of sight passing close to
the edge of the emission cone (impact angle $\beta$ approximately
equal to emission cone radius $\rho$), where the effective emission
beam size is smaller at higher frequencies due to radius to
frequency mapping.

\subsection{Average profiles of \pBn}
\begin{figure*}
\begin{center}
\epsfig{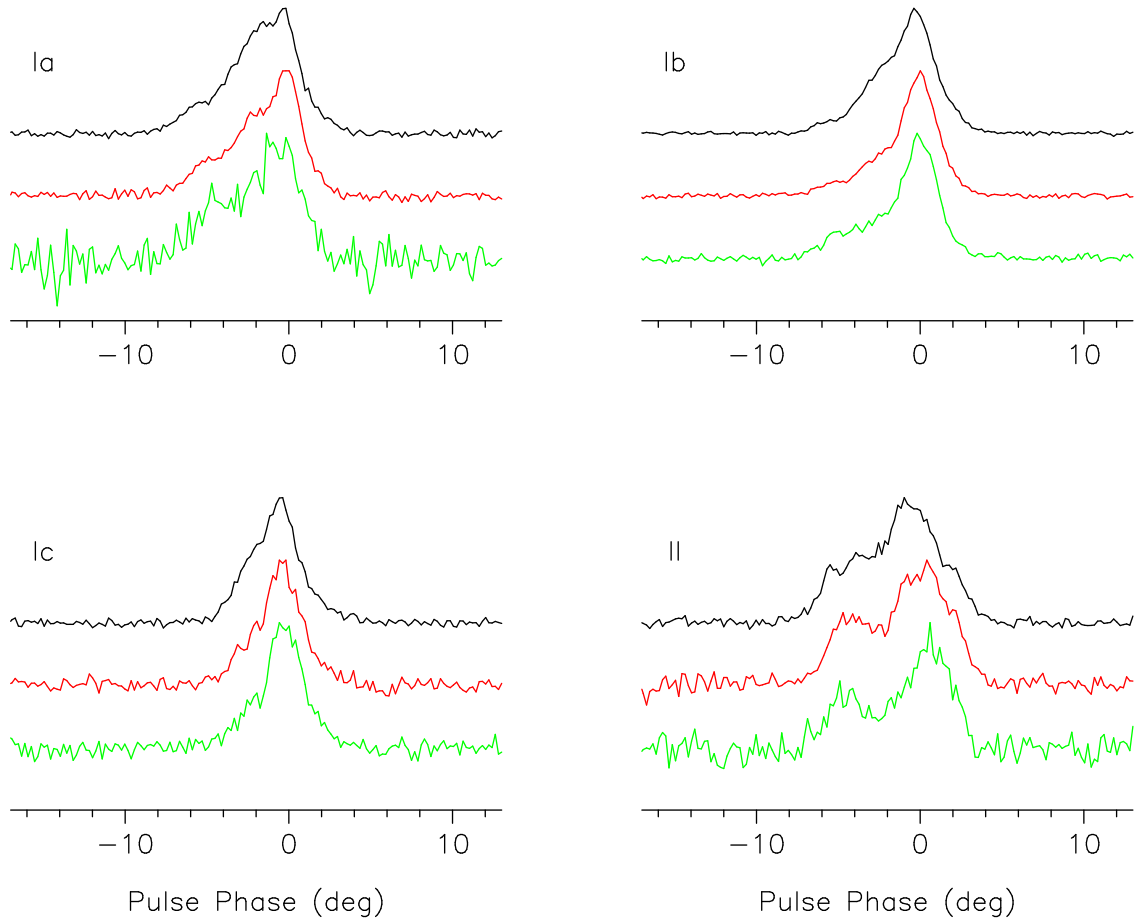}
\caption{Average profiles of \pBn. The four panels (Ia, Ib, Ic and II)
correspond to orbital phase ranges of 171$\deg$-183$\deg$,
183$\deg$-195$\deg$, 195$\deg$-207$\deg$ \& 240$\deg$-273$\deg$,
respectively. The three profiles in each panels are at 820 (top), 1400
(middle) and 2200 MHz (bottom), respectively. Pulse longitude given
in the x-axis has arbitrary reference. See text for details.}
\label{fig:profB}
\end{center}
\end{figure*}

In contrast to \pAn, \pB shows significant change in its pulse profile
structure as a function of orbital phase. For instance, the average
profiles seen in window I are very different from those in window
II. To demonstrate this effect, we have presented the average profiles
derived at three of our bands, 800, 1400 and 2200 MHz, in Figure
\ref{fig:profB}. The orbital phase range corresponding to the four
panels are given in the figure caption. The width of window I
corresponds to approximately 830 seconds ($\sim$35$^{\circ}$ of
orbital phase -- panels $Ia$ to $Ic$). During this interval, the
profile evolves significantly from having a slow rising edge (sharp
trailing edge) to sharp rising edge (slow trailing edge). This is
consistently observed in all the three frequencies. Although there are
some differences in the average profiles in a given orbital phase
range between one orbit to the next, these differences can be
attributed to insufficient averaging in producing the average pulse.

In the fourth panel which corresponds to window II the profile
displays two components, and these two components seem to be well
separated at higher frequencies.

Among classical conal single pulsars, it is common to see profiles
evolving into double profiles at lower frequencies. This is generally
interpreted as due to radius-to-frequency mapping. Also, this model
would certainly predict increasing width of the profiles with
decreasing frequency. However, in the case of panel $(d)$ (window II),
the two components seem to be more pronounced with increasing
frequency, which is the opposite of what we expect from
radius-to-frequency mapping.

Having said this, it is also important to appreciate that in the case
of the classical pulsar population, pulsars are solitary, and there is
no influence on their magnetospheres from the surroundings. However,
as Lyne et al. (2004) state, mechanical luminosity of \pA is
significant enough to influence the magnetospheric properties of
\pBn. In fact, \pBn's proximity to \pA ensures that \pBn's
magnetosphere is altered by \pAn's relativistic plasma wind.

There are three effects that can alter the emission amplitude and
morphology of \pB (Demorest et al.  2004; Arons et al. 2004):
\begin{enumerate}
\item Orbital phase dependence of \pA wind pressure on \pBn's
magnetosphere owing to the orientation of \pAn's magnetic and rotation
axes.
\item Orbital phase dependence of the response of pulsar \pBn's
magnetosphere to the \pA wind pressure, owing to the orientation and
spin phase of \pB as we observe it.
\item Variable expansion of the \pB magnetosphere (at the spin phase
of \pB when we observe \pB) into the downstream cavity created by the
impinging \pA wind.
\end{enumerate}
In all cases the variable magnetosphere shape will likley translate to
variable polar cap size that, in turn, defines the active region of
beamed radio emission which we parameterize with impact angle
($\beta$) and emission cone radius ($\rho$). While the effects in
cases 1 and 2 are likely to produce rather small variations owing to
the likely $P^{1/12}$ dependence of polar cap size on pressure, we
expect that case 3 could lead to substantial variations.

\section{Fluctuation parameters}
\label{sec-fluct}
It is well known that individual pulses from pulsars fluctuate in
their intensity, location and shape, although the average pulse
profiles of almost all pulsars show a great deal of stability. These
fluctuations exhibit a spectacular variety in their spectral
characteristics. The intensity variations at a given pulse longitude
are usually noiselike, although in many cases, there are
quasi-periodic features. These quasi-periodic variations are usually
associated with one of the most remarkable properties seen in pulsar
signals -- drifting subpulses. In this section, we will address
various aspects of single pulse fluctuations seen in signals from \pA
and \pBn.

\subsection{Modulation Index}
\label{sec-modind}
\begin{figure*}
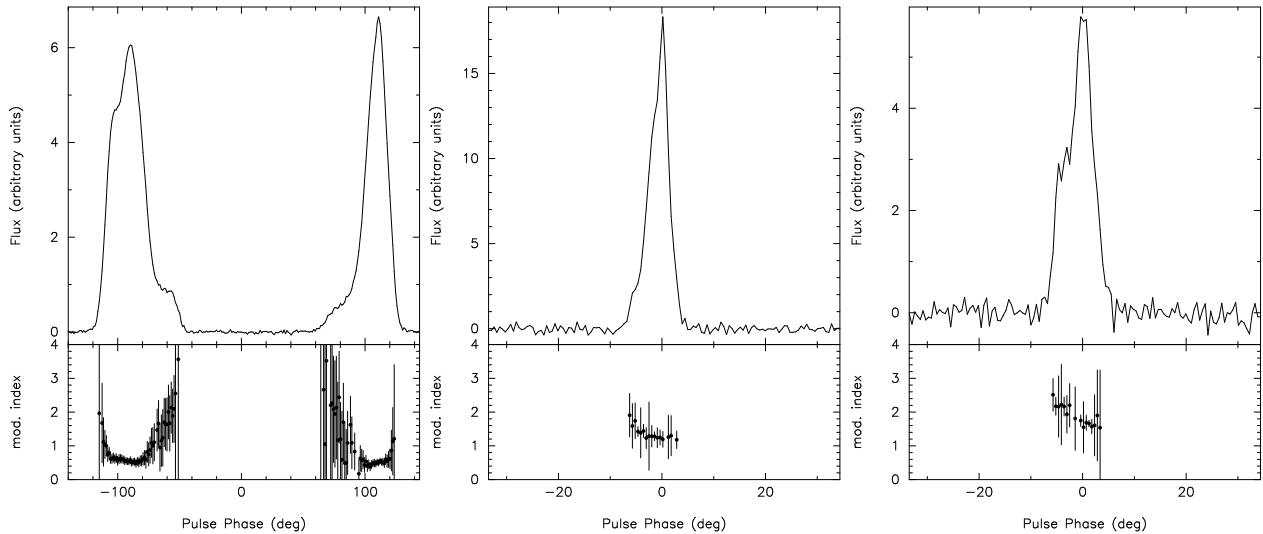

\begin{center}
\epsfig{file=fig4a.ps,height=7cm}
\epsfig{file=fig4b.ps,height=7cm}
\epsfig{file=fig4c.ps,height=7cm}
\caption{Modulation index ($m$) for \pA \& \pBn. ($a$) Upper panel
gives average profile produced with five hundred thousand pulses of
\pAn. Modulation parameter is plotted as a function of pulse longitude
in the lower panel. ($b$) \& ($c$) give the same information for \pB
in window I \& II, respectively. Error bars indicate $1-\sigma$
errors. The longitude resolution in all the plots have been chosen to
optimize signal to noise ratio. See text for details.}
\label{fig:modind}
\end{center}
\end{figure*}

One of the ways to quantify the amount of fluctuations seen in a given
signal is by measuring modulation index, which is measured as
\[
m = \frac{\sqrt{\sigma_{\phi}^2 -
\sigma^2_{ \circ}}}{\mu_{\phi}}, 
\]
\noindent
where $\sigma_{\phi}$ and $\sigma_{\circ}$ are the variance seen in a
given pulse longitude ($\phi$) and the offpulse region, respectively;
$\mu_{\phi}$ is the mean flux measured at that longitude. With our 820
MHz data, for our analysis on \pAn, we considered one hundred thousand
pulses. In Figure \ref{fig:modind}, we give the measured modulation
index as a function of pulse longitude. The left panel corresponds to
\pAn, and the middle and the right panels correspond to windows I and
II of \pBn. The measured value of modulation index comes to almost
hundred percent across the pulse profile. That is, the
r.m.s. variation in the flux level is as high as the mean level
itself.

There are a few investigations reported in the literature that have
measured intrinsic modulation index of pulsars. For instance, Weisberg
et al. (1986) have measured modulation index values of many classical
pulsars. They conclude that core components, on the average, have
lower modulation index than the conal components. While the reason for
this is not completely clear, perhaps it is due to a combination of
intrinsic instability of conal component emissions of pulsars, and
pulse nulling. As Rankin (1986) points out, pulse nulling has {\it
never} been seen before in pulsars with only core emission. There is
also a study of two MSPs by Jenet and his collaborators (Jenet \& Gil
2004; Jenet et al. 2001). They conclude that PSR B1937+21 has very low
-- if not negligible -- intrinsic modulation index, whereas PSR
J0437--4715 exhibits a high modulation index. In general, it is
unclear what the fluctuation properties of MSPs are when compared to
ordinary pulsars, as no comprehensive study exists to date.

Interstellar scintillation is one well known source of extrinsic
modulation that can contribute to the measured value of $m$. The
expected value of $m$ from the interstellar scintillation is given by
$m_{\rm ism} = 1/\sqrt{N_{\rm scint}}$, where $N_{\rm scint}$ is the
number of scintles in the band (Jenet et al. 2001). The observed
modulation index is then given by $\sqrt{m^2+m^2_{\rm ism}}$. The
value of $N_{\rm scint}$ is estimated as $N_{\rm scint}\, = \,(1 \, +
\, \eta B/\Delta\nu)$, where $B$ is the bandwidth of observation,
$\Delta\nu$ is the decorrelation bandwidth, and $\eta$ is the packing
fraction. The measured decorrelation bandwidth of \pA at 800 MHz is
0.1$\pm$0.2 MHz, and at 1400 MHz is 1.8$\pm$0.6 MHz (Ransom et
al. 2004). Given the observed packing fraction of $\sim$0.4, the
estimated modulation index due to interstellar scattering is about 7\%
at 800 MHz. In Figure \ref{fig:modind}, the inner regions of both
features in the \pA profile seems to show significantly higher
modulation index than the outer regions. Such variations in modulation
index as a function of pulse longitude clearly suggests that there is
certainly some intrinsic variation, and not all of it comes from the
interstellar scattering. The high values of modulation index in the
inner regions of the two features also possibly indicate that they are
intermittent.

In windows I and II of \pBn, although modulation index does not vary
much as a function of pulse longitude, it seems to be much higher than
what is expected from interstellar scattering. Therefore, we conclude
that both the pulsars have significant intrinsic variability, whatever
the spectral characteristics may be.

\subsection{Fluctuation properties}
Having established in \S\ref{sec-modind} that \pA and \pB show
significant intrinsic pulse fluctuation, we will assess the spectral
nature of these fluctuations. For instance, these fluctuations could
be almost fully associated with possible drifting subpulses as seen in
many long period pulsars. One of the ways to quantify these
fluctuations is by computing the {\it longitude-resolved} (LR; Backer,
Rankin \& Campbell 1975) and {\it harmonic-resolved} (HR; Deshpande \&
Rankin 1999) spectra.  In general pulsars show wide ranging spectral
properties of pulse-to-pulse fluctuations, from very high coherence
(like in the case of drifting subpulses in PSR B0943+10 \& B0809+74
(Backer, Rankin \& Campbell 1975; Deshpande \& Rankin 2001 and
references therein), or broad quasi-coherent features as in the case
of PSR B2016+28. These (quasi-)coherent features not uncommon among
pulsar population. All `conal-single' pulsars are known to exhibit
drifting subpulse phenomenon (Rankin 1986). Interestingly, highly
coherent fluctuation features are seen in some components of
`multiple-component' pulsars like PSR B1237+25 (Backer, Rankin \&
Campbell 1975; Rankin \& Ramachandran 2003). These coherent features
are thought to represent some highly ordered kinematics in pulsar
magnetosphere.

In the case of \pB, there is also another reason for looking for
possible coherent intermodulations. Jenet \& Ransom (2004) have
recently proposed a scenario to explain the light curve of \pBn. In
their model, the near alignment of the magnetic and the rotation axes
of \A as suggested by Demorest et al. (2004) would mean that the
emission cone of \A is directed towards \B in two phase ranges in the
orbit. Then, if one assumes that \pAn's high energy emission is beamed
the same way as the radio emission, then it is conceivable that the
$\gamma$-ray emission from \pA influences \pBn. They attribute the
bright patches (windows I to IV) of \pBn's light curve to this direct
influence from \pAn.

If \pAn's beamed emission is directly responsible for \pBn's emission,
then it is important to look for possible intermodulation between the
two signals. That is, modulation of \pBn's signal with \pAn's
periodicity. Since the rotation periods of these two pulsars are
incommensurate with each other, if this modulation exists, it will
manifest itself as {\it drifting subpulses} in \pBn's signal. Of
course, this assumes that the environment of \pB does not influence
the beamed radiation from \pA to reduce the effect of
modulation. However, finding such a modulation will be a direct proof
for such an interaction between the two pulsars.

The computed L-R and H-R spectra at 820 MHz and 1400 MHz failed to
show any significant features in windows I \& II, where we had
adequate \snr. Owing to only a limited length of time in each
window, we had only 256 pulses ($\sim$12 min). The corresponding
spectral resolution in the LR spectrum is $(1/256P_1)$, and the
spectral width is $(1/2P_1)$, where $P_1$ is the rotation period.
our sensitivity to detect any fluctuation feature was limited to
one part in 540 in window I, and one part in 220 in window II.

We also looked for possible features in the LR and HR spectra of
\pAn. Although the possibility of having any influence of \pBn's
radiation on \pA is low, \pA may have intrinsic spectral nature of
its fluctuation properties that show (quasi-)coherent features in
the LR and HR spectra. However, our analysis failed to reveal
any such features within our sensitivity limit. Our sensitivity was
one part in $3\times 10^4$, where we considered 500 thousand pulses
for our analysis.

We conclude from these analyses that within our sensitivity limits, we
do not detect any systematic intermodulation of the signals from \pA
and \pBn, and the intrinsic contribution to modulation index comes
from fluctuations of broad spectral nature.

\section{Conclusions}
\label{sec-discuss}
The measured flux from \pB shows systematic variation as a function of
orbital phase. The pulsar is visible is all phase ranges except in
the range of $\phi_{\rm orb}\sim 6\deg$ to $65\deg$, where we have
obtained only an upper limit to the flux. This void region may be
interpreted as an {\it eclipse} of \pB by its own magnetopause region
which is produced by interaction with the relativistic wind from \pAn.
It is this magnetopause region that may also be responsible for the
eclipse of \pA around its superior conjunction (Arons et al. 2004).
Although it is difficult for us to produce reliable light curve around
this \pBn's apparent {\it eclipse} owing to signal-to-noise considerations,
it is clear that the drop in flux is very rapid (within several pulses!)

With our observations at 800 and 1400 MHz, we see some indication of
frequency dependence in \pBn's light curve. During windows III and
IV, the flux at 1400 MHz seems to be systematically lower than the
value at 800 MHz. This needs to be validated with further high
signal-to-noise observations.

Modulation index measurements indicate variation of single pulse flux
in both these objects. These variations are significantly more than
what is expected from the measured decorrelation bandwidth (due to
interstellar scattering) of this system. Therefore, these variations
are intrinsic to the two pulsars. The inner regions in \pA seems to
have significantly higher modulation index than the brighter outer
regions.

Pulsar \A exhibits a complex evolution of its average profile as a
function of radio frequency. However, these properties do not seem to
be different from the characteristics that short period pulsar
population seems to show.

Pulsar \Bn's profile shows dramatic variations as a function of
orbital phase, making it impossible to solve for its geometric
orientation without sensitive polarization measurements. The profile
evolution as a function of frequency seems to reveal mainly a double
component profile, where the components become more pronounced at
higher frequencies. This is quite the opposite of what one would
expect from simple dipole geometry, combined with radius to frequency
mapping.

Given the possible geometrical orientation of \pA derived on the basis
of polarimetric observations, it appears that there will be an orbital
phase dependent wind pressure on \B (Demorest et al. 2004). It is also
plausible that the diameter of \pBn's emission cone changes due to
the effect of orbital phase dependent pressure, there by changing the
fractional value of impact angle ($\beta$) with respect to the radius
of the emission cone ($\rho$). This can introduce complex profile
changes as a function of orbital phase, as we see for \pBn. It is very
important to understand the geometrical orientation of \pBn, for which
sensitive polarimetric observations are necessary.

Within our sensitivity limits, there is no significant intermodulation
between signals from the two objects. We do not detect any significant
feature in the L-R \& H-R spectra, indicating that the spectral
characteristics of the intrinsic pulse fluctuations is
``noiselike''. This constrains the models proposing direct influence
on the emission mechanism of \pB by radiation from \pAn.

\acknowledgements 

We thank the GBT staff, and in particular Carl Bignell, Frank Ghigo,
Glen Langston and Karen O'Neil, for extensive help with the
observations and very useful discussions. V.M.K. is supported from
NSERC Discovery Grant 228738-03, NSERC Steacie Supplement 268264-03, a
Canada Foundation for Innovation New Opportunities Grant, FQRNT Team
and Centre Grants, and the Canadian Institute for Advanced
Research. V.M.K. is a Canada Research Chair and Steacie Fellow.  RR
would like to thank A. A. Deshpande for valuable discussions.

\end{document}